\documentstyle[prl,aps,twocolumn]{revtex}

\begin{document}

\bibliographystyle{prsty}

\title{How to determine Fermi vectors by angle resolved photoemission?}
\author{L. Kipp$^1$, K. Ro{\ss}nagel$^1$, J. Br{\"u}gmann$^1$,
 C. Solterbeck$^2$, T. Strasser$^2$,   W. Schattke$^2$,  M. Skibowski$^1$}
\address{$^1$Institut f\"ur Experimentelle und Angewandte Physik,
 $^2$Institut f\"ur Theoretische Physik und Astrophysik,\\
  Universit\"at Kiel, D--24098 Kiel, Germany}

\author{\parbox{140mm}{\small \vspace{5mm}
~~Angle resolved photoemission spectroscopy (ARPES) has been commonly applied to evaluate the shape of Fermi surfaces by employing simple criteria for the determination of the Fermi vector $\rm k_F$ parallel to the surface such as maximum photoemission intensity at the Fermi level or discontinuity in the momentum distribution function. Here we show that these criteria may lead to large uncertainties in particular for narrow band systems. We develop a reliable method for the determination of Fermi vectors employing high resolution ARPES at different temperatures. The relevance and accuracy of the method is discussed on data of the quasi two dimensional system $\rm TiTe_2$. }}


\maketitle

\vspace{-15mm}

A wide variety of physical phenomena of crystalline materials, 
such as e.~g.\ transport, optical and magnetic response, and 
phase transitions rely on details of the topology of the Fermi 
surface (FS). Its experimental determination  performed by 
traditional techniques like de Haas--van Alphen effect, 
magnetoacoustic effect, Compton scattering, or positron 
annihilation have, however, been restricted to bulk materials. 
All techniques in common is the indirect information on the shape 
of the Fermi surface. While the first two determine extremal 
cross--sections of FS's in a plane normal to the applied magnetic 
field, the latter yield information on one and two dimensional 
projections of FS's. More complex cases, such as superlattices, 
heterostructures or even clean surfaces can also hardly be 
accessed. Angle--resolved photoemission spectroscopy (ARPES) 
has emerged as  probably the most powerful tool for determining 
the occupied electronic band structure of solids and their 
surfaces. Recently, it has been extensively applied to gain 
insight into the topology of Fermi surfaces of a variety of 
materials ranging from conventional three-dimensional metals 
like W and Cu \cite{RotKev98,AvCaAs95,AeOsFa94} to quasi two 
dimensional  layered compounds \cite{StClSt97,Claess98,BoKiSk99}, 
purple bronzes \cite{GwAlCl97}  and high $\rm T_C$ cuprate
 materials \cite{AeOsSc94,SaAvBi97}. The accuracy of the 
 determination of the Fermi surface by ARPES, however, has 
 never been questioned and it turns out that even for the 
 extensively studied BiSrCaCuO it is not at all clear that 
 the topology of the normal state FS shows hole-like pockets 
 around the corners of the Brillouin zone 
 \cite{AeOsSc94,SaAvBi97,DiBeCa96} or electron-pockets around 
 the center of the BZ \cite{ChGrDe99}.

It is widely assumed that ARPES measures the spectral function 
$A({\bf k},\omega)$ of the one-particle system times the Fermi 
function $f(\omega)$ and matrix elements do not play a significant 
role (see e.g.\ ref.\ \cite{NoDiRa98,NoRaDi98}). This motivates 
studies of band dispersions, line shapes, momentum distribution 
functions, and Fermi surfaces. Fermi vectors have been extracted 
from ARPES data employing criteria like (i) maximum ARPES 
intensity at the Fermi level $\rm E_F$ 
\cite{RotKev98,AvCaAs95,GwAlCl97,AeOsSc94,SaTeHi91}, 
(ii) max$\left|\nabla_k\right|$ of the energy integrated 
photoemission intensity \cite{StClSt97,Claess98,CaDiNo96,RoKiFe98} 
or (iii) fitting ARPES peak positions over several emission angles
 and extrapolating the dispersion to $\rm E_F$ \cite{SheDes95}.
However, none of these techniques explicitly considers the detailed 
mechanism of the photoemission process. In particular, matrix 
element effects and differences between photocurrent and spectral
 function have been totally neglected and make these simple 
 interpretations highly questionable. Although, in most cases,
  photoemission peak positions still resemble locations of 
  electronic bands, the shape of spectral functions is in 
  general not directly reproduced by the photocurrent  
  \cite{BoeSch95,SoScZa97}. If photoemission calculations 
  within the one-step model are not available reliable
   simple procedures for an analysis of high resolution
    photoemission data are needed.

Employing high resolution photoemission spectroscopy we 
show in this Letter how Fermi vectors can be determined 
with high accuracy  when intensity modifications due to 
the photoemission process are explicitly eliminated by 
comparing photoemission spectra taken at different temperatures. 
 The reliability and applicability of the method will be 
 demonstrated on the layered Fermi liquid reference material 
 $\rm 1T-TiTe_2$. It does not show an indication of charge 
 density waves and exhibits Ti~3d conduction band emissions 
 which are well separated from other bands 
 \cite{ClAnAl92,HaDuMa94,ClAnGw96}. It may thus serve as a
  model system for assessing the accuracy of the determination
   of the Fermi surface of quasi two dimensional systems.

All results described here were obtained on clean chemical 
vapor transport grown $\rm 1T-TiTe_2$ samples prepared by 
cleavage in ultra high vacuum. Photoemission spectra were 
taken with synchrotron radiation supplied from the storage 
ring DORIS~III at Hamburg Synchrotron Radiation Laboratory 
(HASYLAB) using our Angular Spectrometer for Photoelectrons
 with High Energy REsolution (ASPHERE). The electrons were 
 detected by use of a $180^\circ$ spherical analyzer mounted 
 on a two--axes goniometer with an absolute angular precision 
 of better than $0.1^\circ$. The overall energy resolution 
 was chosen to 30~meV. The position of the Fermi level was 
 determined from spectra of polycrystalline gold with an 
 accuracy of $\rm \pm $1~meV.

The photocurrent within the sudden approximation is given 
by the spectral representation of the one-particle Green 
function: $I({\bf k},\omega) = I_0({\bf k}) f(\omega) 
A({\bf k},\omega)$ where f is the Fermi function and 
$A({\bf k},\omega)$ is the one-particle spectral function. 
The prefactor $I_0$ involves the transition matrix element
 and is thus {\bf k} dependent. As will be shown below this
  {\bf k} dependence crucially affects the reliability of 
  the determination of $\rm{\bf k}_F$. Depending on photon
   energy the component of {\bf k} perpendicular to the
    surface $\rm k_\perp$ is varied in angle resolved
     photoemission. It further complicates a full three     
     dimensional determination of the Fermi vector. This, 
     however, is going beyond the scope of this Letter and 
will be discussed in a forthcoming publication.

The component of the Fermi vector parallel to the surface 
${\bf k}_{F\parallel }$ can be identified by locating the 
peak of $A({\bf k},\omega)$ at zero binding energy. This 
is not a trivial matter since the photocurrent corresponds 
to $f(\omega) A({\bf k},\omega)$ and the peak in the
 photocurrent is what results when the Fermi function 
 cuts off the peak of the spectral function $A({\bf k},\omega)$. 
 Different approaches have been used to solve this difficulty: 
 (i) find the peak in $I({\bf k},E_F)$ (maximum intensity at the 
 Fermi level), (ii)
use the sum rule \cite{RaDiCa95} relating the energy integrated
 spectral function to the momentum distribution function:  
 $n({\bf k}) = \int_{-\infty}^\infty d\omega  f(\omega) A({\bf k},\omega)$. 
 By definition the momentum vectors ${\bf k}_F$ constituting 
 the Fermi surface are identified through a step or at least 
 the vertical slope of the momentum electron distribution 
 $n({\bf k})$ at zero temperature \cite{StClSt97,CaDiNo96}.
Method (iii) employs a fit of photoemission peak positions 
over a wide range of emission angles and extrapolation to 
zero binding energy to obtain an estimate for ${\bf k}_{F\parallel }$.

In Fig.~\ref{simulation} (a) and (b) we show the results of a 
simulation for a narrow band represented by a gaussian peak 
(width: $\rm \Delta E_{FWHM}$~=~100~meV, dispersion: 
$\rm \Delta E / \Delta k$~=~--0.2~eV$\rm \AA$) crossing the
 Fermi level at $\rm k_\parallel=0\AA^{-1}$. Fermi vectors 
 determined using the approaches (i) and (ii)  are plotted 
 as a function of the experimental resolution. Solid lines 
 show results for a matrix element $I_0(k)$ = const, shaded
  areas for a weakly changing  $I_0(k) \sim a(k + 1)$, $-1\le a \le +1$.
Both approaches provide values for $\rm k_{F\parallel }$
 revealing systematic deviations of up to about 20\% 
 (with respect to the distance $\rm \Gamma M=0.96\AA^{-1}$ 
 of the Brillouin zone) strongly depending on the experimental 
 resolution. It should be noted here that these approaches  
 may be additionally affected in strongly correlated Fermi 
 liquid systems where incoherent backgrounds 
 (e.g.\ electron hole pairs) dominate the modulation of 
  $n({\bf k})$ \cite{Matho94}.

A reliable approach for the determination of ${\bf k}_{F\parallel }$
is to compare spectra at different temperatures ($\Delta T$ {\it
method}). Centered at $E_F$ and ${\bf k}_{F}$ and with respect to $\omega$, 
the difference of the Fermi functions for temperatures T$_1$ and T$_2$ 
is an odd function, $A({\bf k}_{F},\omega)$ is even, and therefore the integral of their product
over a symmetric energy window vanishes. In practice, the integration
 is replaced by the finite energy resolution given by a symmetric 
 analyzer function $w$, which vanishes outside 
$[-\epsilon,\epsilon]$. The difference of the intensities taken at
$E_F = 0$ is
 \begin{displaymath} \Delta I({\bf k}_{\parallel}) = I_0 ({\bf k}_{\parallel})
  \int_{-\epsilon}^{\epsilon} d\omega A({\bf k},\omega)
        [f_{T_1}(\omega) - f_{T_2}(\omega)] w(\omega) \; ,
\end{displaymath}
assuming that for the intervals of interest $A$ is independent of
temperature and $I_0$ of temperature and energy.
It then follows from $\Delta I({\bf k}_{\parallel}) = 0$
that ${\bf k}_{\parallel} = {\bf k}_{F\parallel }$. This is  valid for
all $A$ which satisfy in $[-\epsilon,\epsilon]$ that 
$A({\bf k}_F,\omega)=A({\bf k}_F,-\omega)$ and 
$A({\bf k},\omega) \neq A({\bf k},-\omega)$ for ${\bf k} \neq {\bf
k}_F$. Then at general ${\bf k}$ the differences of $A$ do not change sign and 
\begin{displaymath}
\Delta I = I_0
  \int_{-\epsilon}^{0} d\omega (A({\bf k},\omega) - 
                                       A({\bf k},-\omega))
        [f_{T_1} - f_{T_2}] w  \neq 0  \; .
\end{displaymath}
A variety of spectral functions fulfill these conditions,
including Lorentzians, Gaussians, Voigt profiles, 
the Luttinger model \cite{Luttin61}, an extension by Matho 
\cite{HaDuMa94,AlGwCl95}, two dimensional \cite{HoSmWi72} and 
marginal \cite{VaLiSc89} Fermi liquids. A high energy resolution 
extends the scope.
The independence of temperature of such integrals in the vicinity of
${\bf k}_F$ and $E_F$ has already been tested \cite{RaDiCa95} 
and will later be shown again for the present case.
With a high ${\bf k}_{\parallel}$ resolution the criterion is still
valid since for a symmetric band dispersion (with respect to 
($E_F$, ${\bf k}_F$)) the additional contributions annihilate. 
The accuracy might be lowered due to ${\bf k}$ dependencies of 
$I_0$. This possible influence, however, can directly be excluded 
when the experimental $k_\parallel$ resolution windows lies well within
 a regime where point symmetry of $\Delta I(k_\parallel)$ around 
 $\Delta I = 0$ is observed in the experimental data. Note that 
 for a tiny energy interval the required fixed-${\bf k}$ mode of
  photoemission coincides with the employed fixed-angle mode
   well within
the angular resolution. A simulation of the $\Delta T$ {\it method}
 for a gaussian profile and the Luttinger model is shown in 
 Fig.\ \ref{simulation} (c) and (d). The Fermi vector 
 ${\bf k}_{\parallel}=0$ is exactly determined by $\Delta I=0$
  in both cases for all resolutions.

 In Fig.\ \ref{titefig1} (a) we show high energy and 
 angle resolved photoemission spectra of $\rm 1T-TiTe_2$ 
 along the $\rm \Gamma$M direction of the Brillouin zone. 
 For the Ti 3d--band which is well separated from contributions
  of
other bands one observes besides the crossing of the band
 connected
with an intensity break down a characteristic behavior of
 the line shape. Line shape studies have been performed 
 employing either a Fermi liquid scenario using a many 
 body spectral function with an imaginary
part of the self--energy depending quadratically on energy
 referred to $\rm E_F$ \cite{ClAnAl92} or  profiles 
 suggested by Matho \cite{HaDuMa94,AlGwCl95} for 
 approximating the spectral function $A({\bf k},\omega)$. 
 In the close vicinity of the Fermi vector the experimental 
 spectra could excellently be fitted. But for emission 
 angles larger than $\rm \sim 18^\circ$ the quality of 
 the fits is drastically reduced. Values obtained for 
 the Fermi vector were varying between 14$^\circ$ 
 ($\rm 0.51 \AA^{-1}$) \cite{HaDuMa94} and 14.75$^\circ$ 
 ($\rm 0.53 \AA^{-1}$) \cite{ClAnAl92,ClAnGw96,AlGwCl95}. 
 These studies, however, did not consider the photoemission 
 process explicitly and intrinsic spectral information 
 can be distorted by varying matrix elements or other 
 secondary effects. This problem is further illustrated 
 in Fig.\ \ref{titefig1}(a) where an excellent fit 
 (solid lines) can already be achieved for all emission
  angles employing simple Voigt profiles times the Fermi 
  function convoluted by the spectrometer response. This 
  demonstrates that without explicit consideration of the 
  photoemission process sophisticated spectral functions 
  can hardly be delineated. Peak positions resulting from 
  the fit give the experimental dispersion shown in 
  Fig.\ \ref{titefig1} (b). The band crossing is observed 
  at 16.7$^\circ$ corresponding to 0.60~$\rm \AA^{-1}$. 
  It should be noted here that in addition to an increased 
  experimental and calculational (fit) effort the values 
  for $\rm {\bf k}_{F\parallel }$ obtained by method (iii) 
  rely on the profiles employed. For the Matho profile 
  the band crossing is observed at 14.75$^\circ$ 
  (see Fig.\ \ref{titefig1} (b)).

For an application of the $\Delta T$ {\em method}   we plot
normalized photoemission spectra of $\rm 1T-TiTe_2$ taken 
at 100~K and 190~K (Fig.\ \ref{titefig2} (a)). The 
broadening of the Fermi function according to temperature
 is evident. Intensities at the Fermi level 
 ($\rm \Delta E $~=~30~meV) and intensities integrated 
 over the whole spectrum are depicted in (b) and (c), 
 respectively. According to the $\Delta T$ {\em method} 
 the Fermi vector is given by the intersection of the 
 curves in (b) which can clearly be identified at a 
 value of $\rm 16.6^\circ \pm 0.1^\circ$ 
 ($\rm 0.598\pm 0.004$~$\rm \AA^{-1}$). For comparison 
 we also show the values for $\rm k_F$ obtained using 
 the maximum intensity (i) and maximum gradient (ii)
  methods described above. The maximum intensity at 
  $\rm E_F$ can be observed on a relatively broad peak 
  at around 17.8$^\circ$ ($\rm 0.64 \AA^{-1}$) showing 
  the systematic erroneous shift of $\rm k_{F\parallel}$
   towards occupied states  as already observed in the
    simulation of Fig.\ \ref{simulation}. The energy 
integrated intensity and its derivative are depicted in 
Fig.\ \ref{titefig2} (c). Since the gradient of the 
integrated intensity is only marginally 
changing in the $\rm k_\parallel$ regime of interest a 
rather broad maximum is observed making a detailed 
quantitative comparison with FS calculations very 
difficult. 
Symmetrized spectra $I({\bf k_F},\omega)+I({\bf k_F},-\omega) 
= I_0({\bf k_F})A({\bf k_F},\omega)$ eliminating the Fermi 
function \cite{assume,NoDiRa98} are shown in Fig.\ \ref{titefig2} 
(d). For the spectra taken at 100~K and 190~K no differences 
can be observed ruling out any T-dependence of 
$I_0({\bf k_F})A({\bf k_F},\omega)$ in this temperature range.

In summary, we have developed a reliable and simple method 
to determine Fermi vectors by angle resolved photoemission 
spectroscopy. Employing temperature difference spectra the 
photoemission process hiding the spectral function in 
measured spectra is explicitly considered. Therefore, 
the method even works for systems showing rapidly 
changing matrix elements with variation of $\bf k$. 
For the layered material $\rm 1T-TiTe_2$ we have 
demonstrated that the accuracy of the determined 
$\rm k_F$ values can be significantly improved to 
better than $\pm$0.4\% of the dimension of the Brillouin zone. 
Compared to other methods revealing up to one order of 
magnitude larger error bars together with systematic 
deviations of up to 20\% reliable quantitative comparisons 
between experiment and Fermi surface calculations going 
beyond {\em similarities of the shapes} will now 
become possible.\\

Experimental help by R. Schwedhelm and S. Woedtke is gratefully 
acknowlegded.
This research is supported by the BMBF, FR Germany
(project No. 05~SB8~FKB, 05~SE8~FKA and 05 SB8 FKA).
~

\begin{figure}[t]
  \caption[simulation]{\label{simulation} Systematic deviations 
  for $\rm k_{F\parallel }$ determined by experimental procedures
   employing  methods (i) $\max(I({\bf k},E_F))$ and (ii) 
   $\max(\left|\nabla_k n(k)\right|)$ plotted as a function 
   of experimental resolution (gaussian type) for 300~K (a) and 30~K (b). 
   Simulation for a gaussian peak (width: $\rm \Delta 
   E_{FWHM}$~=~100~meV, dispersion: 
   $\rm \Delta E / \Delta k$~=~--0.2~eV$\rm \AA$) crossing 
   the Fermi level at $k_{F\parallel }=0$~$\rm \AA^{-1}$. 
   Solid lines and shaded areas represent simulations for 
    constant and varying matrix elements, respectively. 
    Simulation results for the $\Delta T$ {\em method} are
     shown in (c) and (d). See text for details.} 
\end{figure}

\begin{figure}[t]
  \caption[titefig1]{\label{titefig1} (a) Angle resolved 
  photoemission spectra (EDC)
		associated with  the
		Ti 3d--band of $\rm 1T-TiTe_2$ in the
		 direction $\Gamma$M taken with high
		resolution and line shape fit using Voigt
		 profiles. (b) Band dispersion obtained 
		 from the fitted peak positions (method 
		 (iii)) using Voigt and Matho profiles. 
		 Note the different band crossings at $\rm E_F$.} 
\end{figure}

\begin{figure}[t]
  \caption[titefig2]{\label{titefig2}(a) Angle resolved 
  photoemission spectra ($\rm \Delta E$~= 30~meV) associated 
  with  the
		Ti 3d--band of $\rm 1T-TiTe_2$ in the 
		direction $\Gamma$M taken at 100~K 
		and 190~K. (b) Intensities at the Fermi 
		level (method (i)). The intersection
		 marks the Fermi vector ($\Delta T$ 
		 {\em method}). An uncertainty of 
		 $\pm 0.1^\circ$ (hatched area) 
		 emerges from the error bar of 
		 the Fermi energy ($\rm \pm 1$~meV) . 
		 (c) Intensities integrated over the
		  whole spectrum and k--derivative 
		  (smoothed) of the integrated intensity 
		  (method (ii)). Grey bars mark the 
		  uncertainties for the determination 
		  of $\rm k_F$ (95\% of the maximum value). 
		  (d) Symmetrized spectra showing no 
		  differences for spectral functions 
		  at 100~K and 190~K. } 
\end{figure}

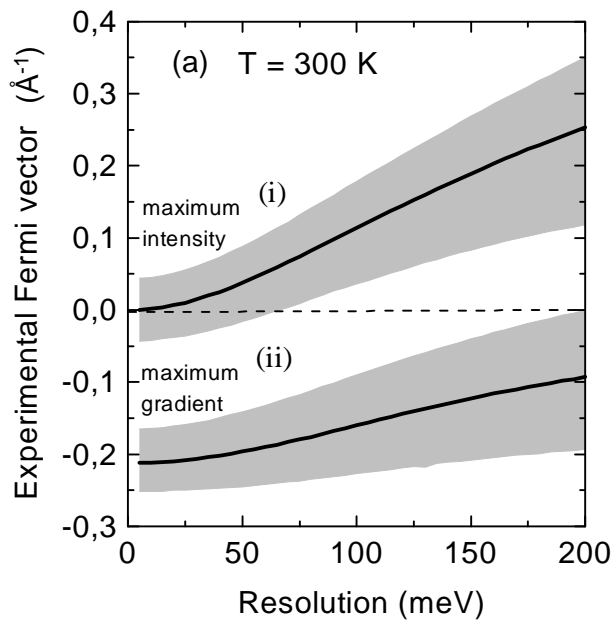
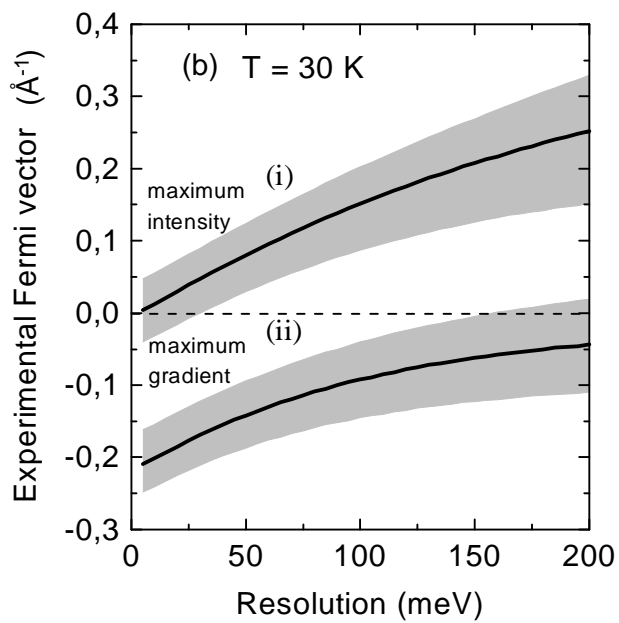
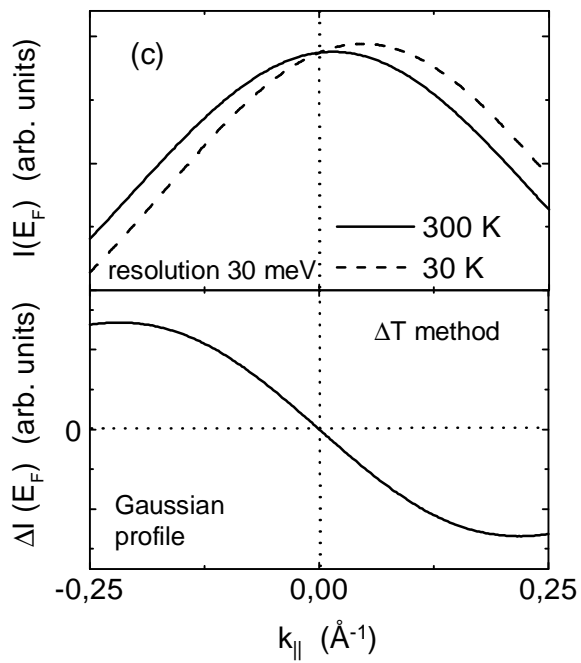
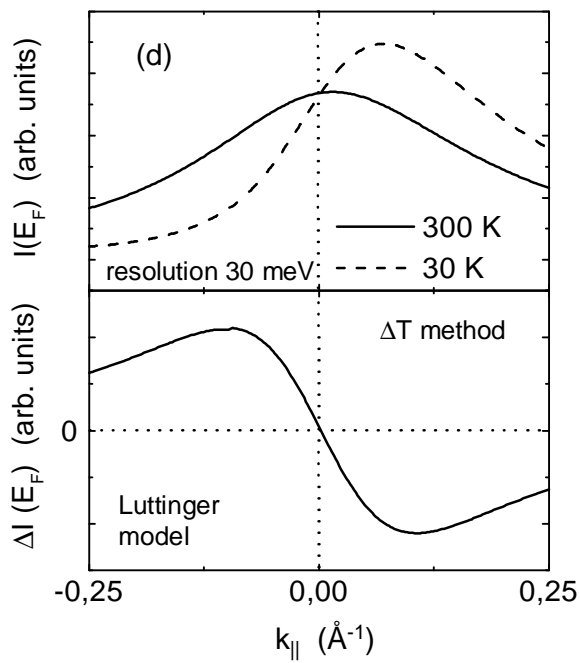
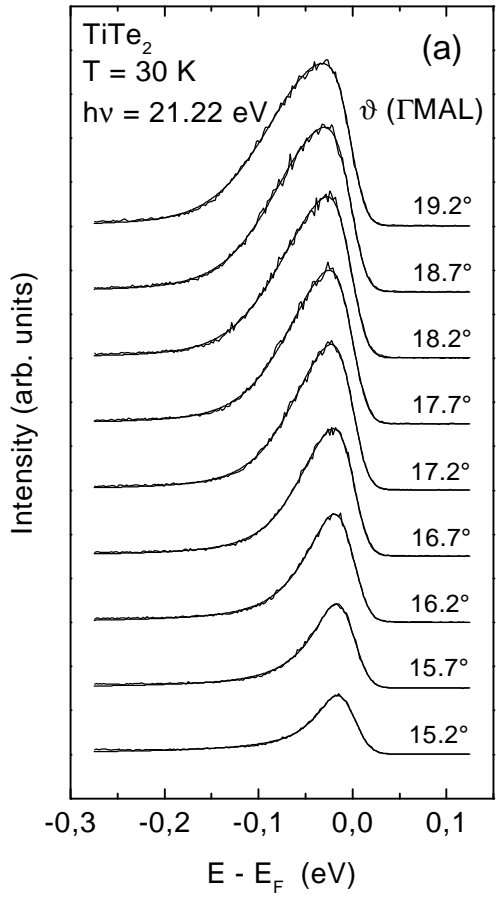
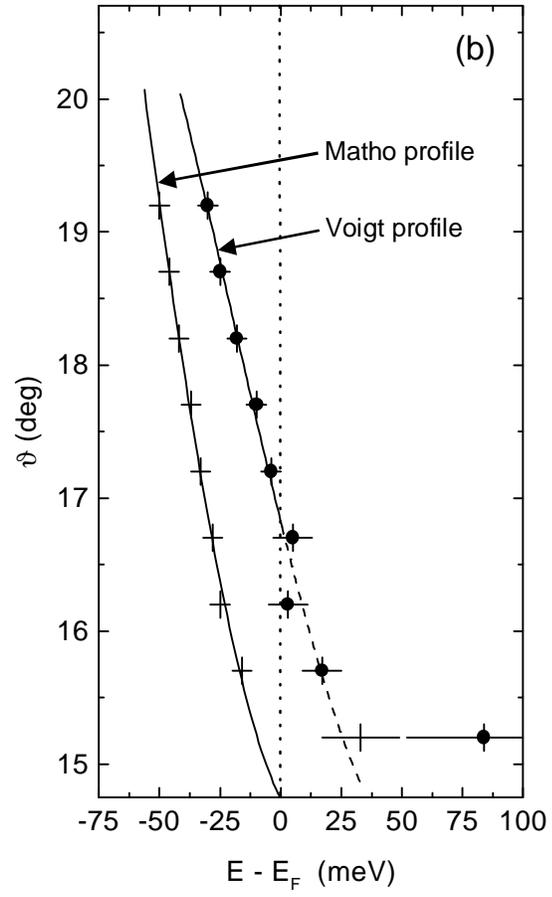
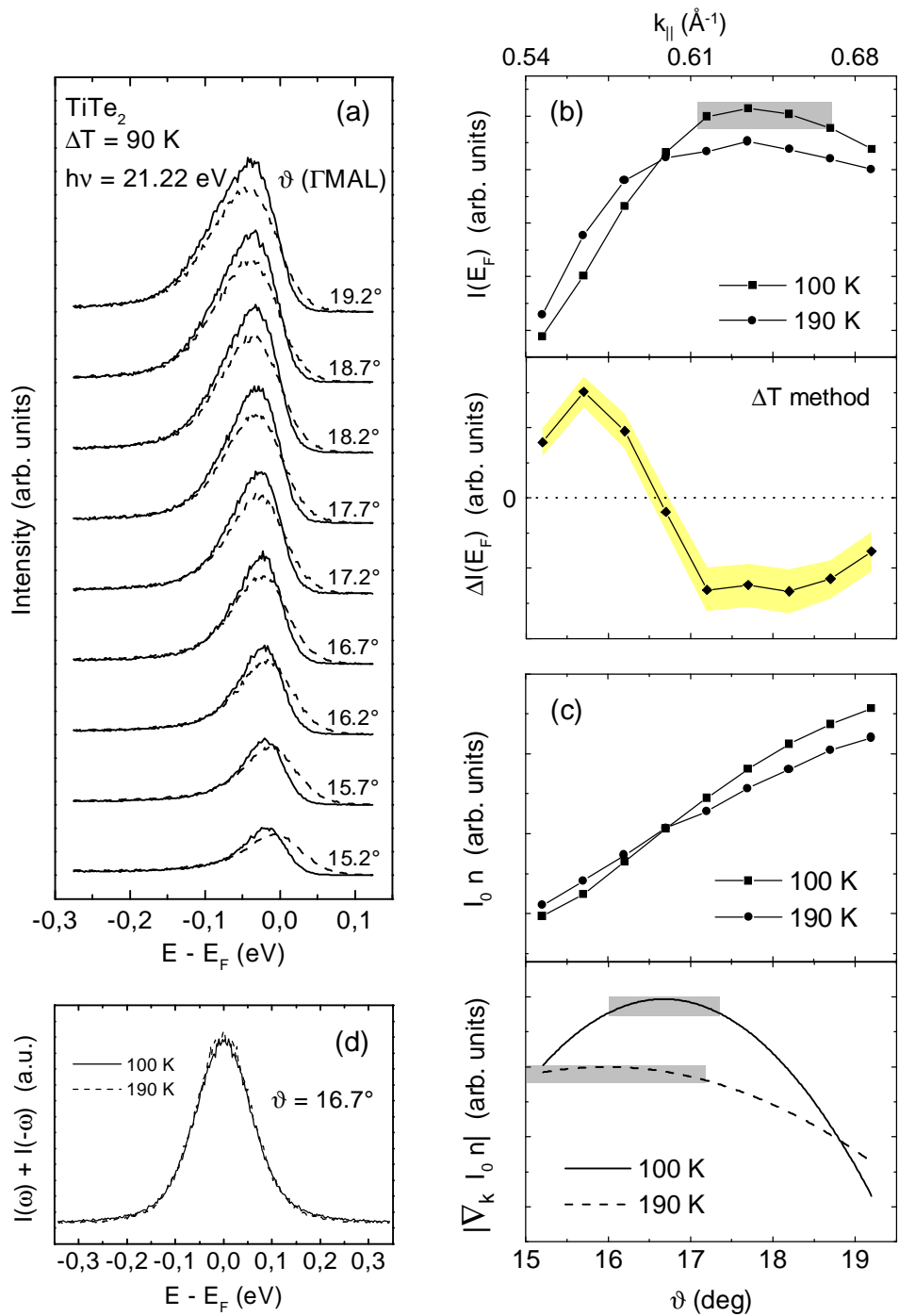
\end{document}